\documentclass[prl,twocolumn,showpacs,preprintnumbers,twoside]{revtex4}

\usepackage{graphicx}

\newcommand\ee{\end{equation}}
\newcommand\be{\begin{equation}}
\newcommand\eea{\end{eqnarray}}
\newcommand\bea{\begin{eqnarray}}

\newcommand\hogw{h _0 ^2 \Omega_{\rm gw}}
\newcommand\hocoll{h _0 ^2 \Omega_{\rm coll}}

\begin{document}

\title{Relic gravitational waves from colliding bubbles and cosmic turbulence}
\author{Alberto Nicolis}
\affiliation{Instituto de Estructura de la Materia, CSIC, Serrano 123, 28006 Madrid, Spain}
\begin{abstract}
A stochastic background of gravitational waves can be generated during a cosmological 
first order phase transition, 
at least by two distinct mechanisms: collisions of true vacuum bubbles and turbulence 
in the cosmic fluid. 
I compare these two contributions, analyzing their relative importance for a generic phase transition.
In particular, a first order electroweak phase transition is expected to generate a gravitational 
wave signal peaked at a frequency which today falls just within the band of the planned space 
interferometer LISA.
For this transition, I find constraints for the relevant parameters 
in order to produce a signal within the reach of the sensitivity of LISA. The result is that
the transition must be strongly first order, $\alpha \gtrsim 0.2$. In this regime the
signal coming from turbulence dominates over that from colliding bubbles.
\end{abstract} 
\date{March, 2003}
\preprint{IEM-FT/230-03}
\preprint{gr-qc/0303084}
\maketitle

Several phase transitions have probably taken place during the evolution of the Universe, and
among the possible remnants of these cosmic events there is a stochastic background
of gravitational waves (GWs).
During a first order phase transition the Universe finds itself in a metastable phase, the
false vacuum, separated from the true vacuum by a potential barrier in the free energy of an
order parameter, usually a scalar field $\phi$. Quantum tunneling throughout the barrier can take place
in finite-volume regions of space, thus giving rise to the nucleation of true vacuum bubbles inside
the false vacuum phase. 
As the Universe expands and the 
cosmic temperature $T$ drops down, the energy difference between the false and the true vacuum
gets larger, and correspondingly 
the phase transition gets more and more probable. 
Once the probability of nucleating one critical bubble per Hubble volume per Hubble time 
becomes ${\cal{O}}(1)$ the transition begins. This implicitly defines
the transition temperature $T_*$, by the relation
$\Gamma (T_*) \sim H_*^4$, where $\Gamma$ is the critical bubble nucleation rate 
per unit volume, and $H_*$ is the Hubble parameter. 
The nucleated bubbles begin to expand and collide, eventually
filling up the whole Universe. When two or more bubbles collide spherical symmetry is
broken, and a small fraction of their kinetic energy is released in form of GWs.
Moreover, since the bubbles expansion is accompanied by macroscopic motions in the cosmic plasma, 
the collision of two (or more) bubbles results in some sort of anistropic stirring of the plasma.
Given the high Reynolds number at such high cosmic temperatures, turbulent motions develop, thus
providing another potentially interesting source of GWs.

The GW background produced by bubble collisions in a first order phase transition 
has been extensively studied in the literature by means of analytical estimates 
and numerical simulations
\cite{witten, hogan,thorne,TW,KTW,KT,KKT}.
On the other hand, only very recently a systematic computation of the amount of GWs emitted
by turbulent fluid motions has been performed \cite{KMK,DGN,tesi}, showing that turbulence can be
indeed a powerful source of GWs, even more effective than the bubble collision process.
The total stochastic GW signal one can expect from a generic first order transition
is the sum of these two contributions: the goal of the present paper is to compare them, and
to find under which conditions one dominates over the other. Moreover,
the electroweak phase transition (EWPT), if it is of first order, is expected to give rise
to a GW background peaked around the milliHertz frequency, and this happens to be in the range
most relevant for the space interferometer LISA \cite{LISA}, which is planned to be launched
by 2011. I will find what constraints must be satisfied by the EWPT in order to give
a GW signal within the sensitivity of LISA.

\begin{figure}
\begin{center}
\includegraphics[width=7cm]{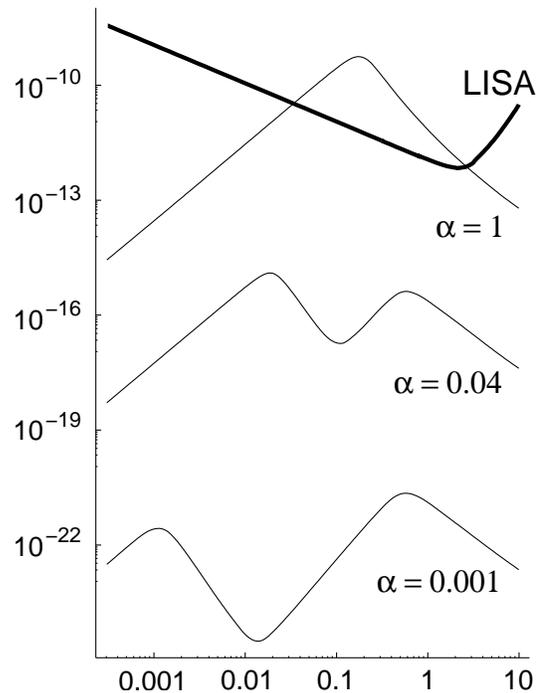}
\caption{\label{spettri_fig} The total GW spectrum $\hogw$ {\em vs.}~frequency (in mHz),
	from a generic first order phase transition,
	for three different values of $\alpha$, together with the sensitivity curve of LISA (bold line)
	\cite{LISA}.
	$\beta / H_*$ is set to 100 and $T_*$ to 100 GeV: different values would
	only give a global rescaling of the spectra, without affecting their shape.}
\end{center}
\end{figure}

An useful parametrization of a stochastic background of GWs \cite{thorne, maggiore} 
is given by the dimensionless quantity
\be
\Omega_{\rm gw} (f) \equiv \frac 1 {\rho_{\rm c}} \frac{d \rho_{\rm gw}}{d \log
f}      \; ,
\ee
which is the GW energy density per frequency octave, normalized to the critical
energy density
for closing the Universe, ${\rho_{\rm c}} = {3 H_0 ^2}/{(8\pi G)}$,
with $H_0 = h_0 \times 100 \mbox{ km} / ({\rm sec} \cdot {\rm Mpc})$; $h_0$ parametrizes
the experimental uncertainty in $H_0$. In fact, it is more convenient to characterize the 
GW energy spectrum with $\hogw (f)$, which is independent on $h_0$.
The LISA sensitivity to $\hogw$ is shown in fig.~\ref{spettri_fig}.

For the purposes of the present paper, we can summarize the dynamics of a first order phase
transition into
two quantities, traditionally called $\alpha$ and $\beta$.
The former, $\alpha$, is defined as the ratio between the false vacuum energy (latent heat) 
density and the plasma thermal energy density, both computed at the transition temperature $T_*$. 
It gives a measure of the transition strength: for $\alpha \ll 1$ the energy jump experienced
by the system in performing the transition is neglectable with respect to the average thermal energy,
and thus the transition is very weak; if on the contrary $\alpha \sim {\rm few}$, the transition is
very strongly first order.
The latter quantity, $\beta$, is the rate of (time) variation of the nucleation rate 
itself, computed at the transition time $t_*$, $\beta \equiv \dot \Gamma / \Gamma |_*$.
Therefore $\beta^{-1}$ fixes the typical time scale of the phase transition: once the transition
has begun, in a time interval $\Delta t \sim \beta^{-1}$ the whole Universe is
converted to the true vacuum phase. 
This is because typically 
$\beta$ is much larger than the Hubble parameter at the transition time, $\beta \gg H_*$
\cite{TWW}.
In spite of the simplicity of their definition,
the determination  of $\alpha$ and $\beta$ for a particular first order phase transition, 
once a specific particle physics model is given,
requires involved numerical computations (see for instance ref.~\cite{AMNR}).

Two more quantities are important in determining the amount of GWs emitted by bubble collisions
and turbulent motions. These are 
$\kappa$, the fraction of vacuum energy (latent heat) that is transformed into fluid 
kinetic energy, rather than into heat, and $v_b$, the velocity at which the bubble walls expand.
For a general phase transition their determination involves taking into account frictive effects
due to departures from thermal equilibrium
in the vicinity of the wall, 
which is difficult \cite{turok,MP,JS}. 
However, in the case of a {\em detonation} -- {\em i.e.}~if 
the bubble wall propagates faster than the speed of sound ($1/\sqrt{3}$, in a 
relativistic thermal bath); the opposite case is called {\em deflagration} --
both $v_b$ and $\kappa$ are functions of $\alpha$ only,
independently on the particular microphysics that drives the transition.
Throughout this paper we consider only detonations, for the following reasons.
First, it has been argued \cite{KF} that the detonation mode is stable against nonspherical
hydrodynamical pertubations, while deflagration is not. 
Second, and more important from our perspective, in a 
deflagration there is no large concentration of kinetic energy near
the bubble wall, characteristic velocities are small, and thus GW production 
is strongly suppressed. 
We will show that GWs produced at the EWPT
will be detectable at LISA only if the phase transition is very strong, 
and in such case the bubble expansion proceeds via detonation \cite{KKT}.
For a detonation the expansion velocity $v_b$ turns out to be \cite{stein}
\be	\label{vb}
v_b (\alpha) = \frac{1/\sqrt{3}+\sqrt{\alpha^2+2 \alpha/3}}{1+\alpha}	\;.
\ee
It is an increasing function of $\alpha$ and varies between the speed of sound 
$1/\sqrt{3}$ and the speed of light.
Similarly, a numerical computation gives \cite{KKT}
\be	\label{kappa} 
\kappa (\alpha) \simeq \frac 1{1 + 0.715 \: \alpha} \left[0.715 \: \alpha + 
	\frac{4}{27} \sqrt{\frac{3\alpha}{2}}	\right] 	\; ,
\ee
which shows that also the efficiency $\kappa$ is increasing with increasing $\alpha$, varying 
between 0 and 1.

The energy spectrum of GWs produced by bubble collisions is peaked at a frequency of 
about $2 \beta$, independently on the bubble
expansion velocity $v_b$ \cite{KKT}. 
This shows that $\beta^{-1}$ is indeed the relevant time scale of
the process. 
Redshifting that frequency till the present time gives
\be	\label{f_coll}
f_{\rm coll} \simeq 5.2 \times 10^{-3} \: {\rm mHz} \: \left[ \frac{\beta}{H_*}\right]
	\left[ \frac{T_*}{100 \: \rm GeV}\right] \left[ \frac{g_*}{100}\right]^{1/6}	\; ,
\ee
where $g_*$ is the number of relativistic degrees of freedom at the transition time.
Typical values for $\beta / H_*$ for the 
EWPT are  between $10^2$ and a few times $10^3$, with $T_* \sim 100$~GeV. 
This gives a frequency $f_{\rm coll}$ in the range $(0.5 - 10)$ mHz, which 
is precisely the range in which LISA achieves its maximum sensitivity.
The present time fraction of critical density stored in GWs from bubble 
collisions at the peak frequency $f_{\rm coll}$ is \cite{KKT}
\bea	
\hocoll (f_{\rm coll}) & \simeq & 1.1 \times 10^{-6} \: \kappa^2 \left[\frac{H_*}{\beta}\right]^2 
\left[ \frac{\alpha}{1+\alpha} \right]^2 \times \nonumber \\ 
&& \times \left[ \frac{v_b ^3}{0.24+v_b^3} \right]
\left[ \frac{100}{g_*}\right]^{1/3}	\label{omega_coll}		\; .	
\eea
At low frequencies the spectrum $\hocoll (f)$ increases as $f^{2.8}$, while
at high frequencies it drops off as $f^{-1.8}$ \cite{KTW}.

When bubbles collide, the plasma is stirred up at a length scale comparable to the size
of the colliding bubbles. Larger bubbles are more energetic than smaller ones, and
indeed it can be shown that the most part of the kinetic energy involved in the process 
is released at the largest length scale in the system, namely the radius of the largest bubbles
at the end of the transition \footnote{This characteristic length turns out to be five 
times larger than the naive expectation $\beta^{-1} v_b$.
This leads to an enhancement of the signal, for instance with respect to the results 
of ref.~\cite{KMK}, see ref.~\cite{DGN} for details.
}.
This stirring process creates large eddies in the plasma, with a characteristic turnover velocity
$u_S$ given by \cite{tesi}
\be	\label{u_S}
u_S \simeq \sqrt{\frac{\kappa \: \alpha}{\frac 4 3 + \kappa \: \alpha}}	\; . 
\ee
Note that, if the phase transition proceeds through detonation bubbles (as we are assuming), 
also $u_S$ can be expressed
uniquely in terms of $\alpha$, by means of eq.~(\ref{kappa}).
Moreover, in such a case it turns out that $u_S$ is always smaller than the bubble 
wall velocity $v_b$, eq.~(\ref{vb}), and they both approach the speed of light 
in the strong transition limit, $\alpha \sim {\rm few}$.
Once the large eddies have formed, after a few revolutions they decay into smaller ones,
thus giving rise to the usual turbulent energy cascade. The turbulent motions that 
develop source a stochastic backgroud of GWs peaked at a present ({\em i.e.}~properly
redshifted) frequency \cite{DGN}
\be     \label{f_turb}
f_{\rm turb}  \simeq 3.4 \times 10^{-3} \: {\rm mHz} \:
\frac{u_S}{ v_b} \: 
\left[ \frac{\beta}{H_*} \right]
\left[ \frac{T_*}{100 \: \rm GeV} \right] \left[ \frac{g_*}{100} \right] ^{1/6}
\; .
\ee 
If the transition is strong enough $u_S / v_b \sim 1$ and thus $f_{\rm turb} \sim f_{\rm coll}$.

The relic GW energy density coming from turbulence, computed at the peak frequency $f_{\rm turb}$, is 
\cite{DGN,tesi}
\be	\label{omega_turb}
h_0 ^2 \Omega_{\rm turb} (f_{\rm turb}) \simeq 1.4 \times
10^{-4}
\:
u_S ^{5} \: v_b ^{2}
\left[ \frac{H_*}{\beta}  \right]^2
\left[ \frac{100}{g_*} \right] ^{1/3}
\; .
\ee
At frequencies larger than $f_{\rm turb}$ the GW energy spectrum decreases as $f^{-7/2}$,
while at lower frequecies it rises as $f^2$.
Strictly speaking, these results holds if magnetohydrodynamical (MHD) effects are absent
or negligible during the phase transition considered, so that the cosmic fluid develops 
standard Navier-Stokes
turbulent motions, with an ordinary Kolmogorov energy spectrum.
This approach, however, is not always justified: for instance, large magnetic fields may have
been produced before or during the EWPT by a number of different processes \cite{report,Baym}. 
The inclusion of MHD effects in the above GW spectrum is
straightforward: one has simply to replace the high frequency behaviour $f^{-7/2}$ 
with $f^{-8/3}$, the peak frequency $f_{\rm turb}$ and the peak value of $h_0 ^2 \Omega_{\rm turb}$
being unaltered \cite{DGN}.

Since in a phase transition both bubble collisions and turbulent motions may be
powerful sources of GWs, it is interesting to compare characteristic frequencies and
energy spectra of the GW backgrounds from the two processes, it order to understand what is 
the dominant contribution to a possibly detectable signal. 

First, note that the expected peak frequency of the GW background from bubble collisions, 
eq.~(\ref{f_coll}), is
always larger than that of the signal coming from turbulent motions, 
eq.~(\ref{f_turb}), although in the strong transition limit $\alpha \gtrsim 1 $ they become
of the same order. 
Their ratio is indeed
\be
\frac{f_{\rm turb}}{f_{\rm coll}} \simeq 0.65 \: \frac{u_S}{v_b}	\; ;
\ee
if the transition proceeds through detonation bubbles, this ratio is a function of $\alpha$ only.

Then, we compare the relic energy amounts.
Given their different behaviors with frequency, the dominant contribution comes from bubble collisions 
at high frequencies and from turbulence at low frequencies.
The qualitative picture we can expect at intermediate frequencies is the following.
In moving from lower to larger frequencies, we first encounter the turbulence peak at $f_{\rm turb}$, 
then the spectrum drops down and, when we reach $f_{\rm coll}$, we encounter the bubble-collisions peak.
It may happen then the second peak is overwhelmed by the 
high frequency tail of the turbulence signal.
As we show below, this possibility depends on the value of $\alpha$.
However, even if this is the case, at larger frequencies the contribution from 
bubble-collision is going to become dominant, thus giving rise to a change in the slope of the 
total GW spectrum.
The opposite case, namely that the turbulence peak is overwhelmed by the low frequency tail coming
from bubble collisions, is in principle possible, but it is quantitatively ruled out for
any value of the parameters: the turbulence peak height is always much larger than
the height of the low frequency tail from bubble collisions computed at $f_{\rm turb}$.
Fig.~\ref{spettri_fig} schematically shows the total GW spectrum for different values of $\alpha$.

We can define two interesting quantities.
The first is the ratio of the GW energy spectra, each computed at its own peak frequency,
\be
r_1 \equiv \frac{\Omega_{\rm turb} |_{f_{\rm turb}}}{\Omega_{\rm coll} |_{f_{\rm coll}}} \simeq
	130 \: u_S^5 \: \frac{1}{\kappa^2} \: \frac{0.24+v_b^3}{v_b}\left[ 
	\frac{1+\alpha}{\alpha}\right]^2   \; .
\ee
The second is the same ratio, but with both the spectra computed at the bubble collision peak frequency;
for Navier-Stokes turbulence this is
\be
r_2 ^{\rm NS} \equiv \left. \frac{\Omega_{\rm turb} ^{\rm NS}}{\Omega_{\rm coll}}
	\right| _{f_{\rm coll}} \simeq
	30 \: u_S^{17/2} \: \frac{1}{\kappa^2} \: \frac{0.24+v_b^3}{v_b^{9/2}}
	\left[ \frac{1+\alpha}{\alpha}\right]^2   	\; ,
\ee
while for MHD turbulence it becomes
\be
r_2 ^{\rm MHD} \equiv \left. \frac{\Omega_{\rm turb} ^{\rm MHD}}{\Omega_{\rm coll}}
	\right| _{f_{\rm coll}}  \simeq
	50 \: u_S^{23/3} \: \frac{1}{\kappa^2} \: \frac{0.24+v_b^3}{v_b^{11/3}}
	\left[ \frac{1+\alpha}{\alpha}\right]^2   \; .
\ee
In the case of bubbles expanding as detonation fronts, both $r_1$ and $r_2$ depend only on $\alpha$.
In particular, they are increasing functions of $\alpha$, they go to zero in the weak transition limit,
$\alpha \to 0$, and get very large 
in the strong transition limit, $\alpha \gtrsim 1$.
Therefore, if the phase transition is very strongly first-order the bubble-collision peak is
overwhelmed by the signal coming from turbulence, although at higher frequencies
it is the bubble-collision contribution that dominates the spectrum. This is the top case in 
fig.~\ref{spettri_fig}, and happens if $r_2>1$, which in terms of $\alpha$ reads $\alpha>0.67$ in
the Navier-Stokes turbulence case and $\alpha>0.44$ in the MHD one.
For smaller values of $\alpha$ the spectrum shows both the peaks: the bubble-collision one is higher
than the turbulence one if $r_1 < 1$, or equivalently $\alpha < 0.012$. This is the bottom case in the
figure, and corresponds to a weak transition.
In the intermediate case $0.012 < \alpha < 0.67 \:(0.44)$ the turbulence peak is
higher than the bubble-collision one, see the center plot in the figure.

\begin{figure}[b]
\begin{center}
\includegraphics[width=7cm]{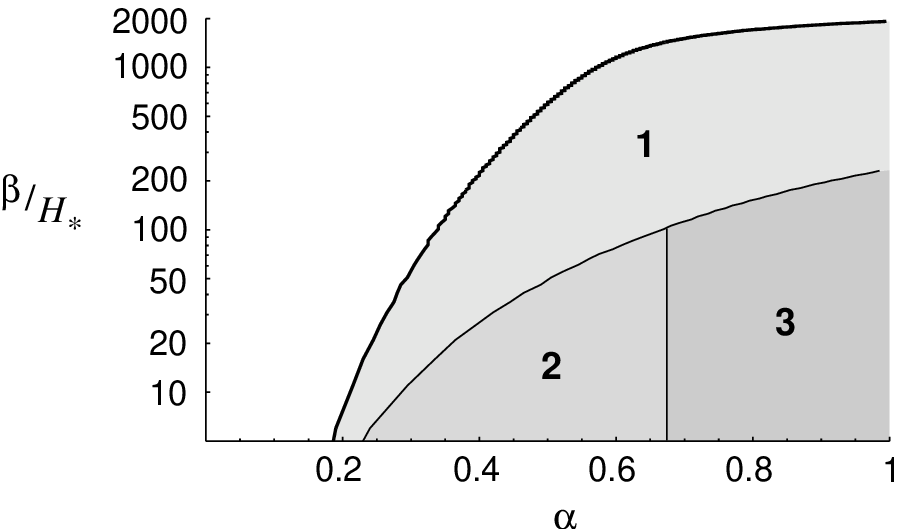}	\\ 
\includegraphics[width=7cm]{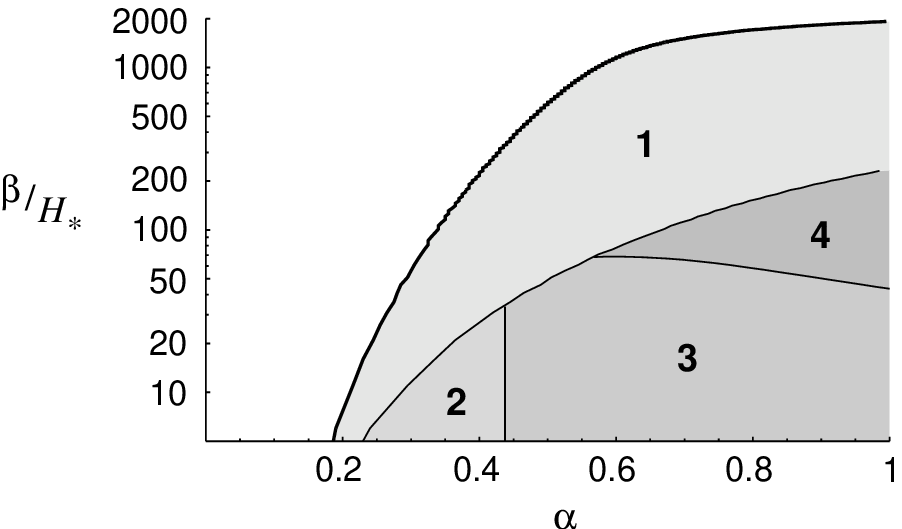}
\caption{\label{turbo_vs_bubble} The shaded regions show the values of $\alpha$ and $\beta/H_*$
	for which the total GW signal coming from a first order electroweak phase transition is
	above the LISA sensitivity curve. $T_*$ is fixed to 100 GeV.
	The top plot refers to ordinary Navier Stokes turbulence, while the bottom one 
	to MHD turbulence.
	In the region labelled by {\bf\sf 1} only GWs from turbulence are visible at LISA.
	In {\bf\sf 2} both peaks (from turbulence and from bubble collision) are visible. 
	In {\bf\sf 3} the bubble collisions peak is below the high frequency tail 
	coming from turbulence, but nevertheless there exists a frequency range in which 
	the bubble collision contribution is recognizable by the slope change in the signal.
	In {\bf\sf 4} (in the bottom plot) such slope change takes place at too high frequecies,
	where the GW signal is below the LISA sensitivity, and therefore, the bubble collision
	signal is completely overwhelmed by the turbulence one in the frequency window interesting
	for LISA.}
\end{center}
\end{figure}

Finally, we compare the expected GW signals from bubble collisions 
and from turbulence with the LISA sensitivity curve shown in
fig.~\ref{spettri_fig}. 
This easily permits to find the values of $\alpha$ and $\beta$ for which the GWs coming
from a (possibly) first order EWPT are detectable at LISA.
To do that, we fix $T_*$ at 100 GeV \footnote{The actual value of $T_*$ affects the 
peak frequency, but not the energy spectrum at the peak. However,
for the electroweak transition the transition temperature is always close to 100 GeV.}
and scan the $(\alpha, \beta)$ plane. For each point considered we superimpose the total GW spectrum
on the LISA sensitivity curve, and check if there is a range of frequencies for which the signal
is above the instrumental sensitivity and, in such a case,
which is the background really detectable, either the one from turbulence or the one 
from bubble collisions, or both.
The results of such a procedure are shown in fig.~\ref{turbo_vs_bubble} and described in
the caption.
The conclusion we can draw from the figure is that in order to generate a GW signal within the reach of
the sensitivty of LISA the electroweak phase transition must be strongly first order, 
$\alpha \gtrsim 0.2$. In this regime the contribution from turbulence to the total GW signal is likely
to be more important than that from bubble collisions.
Such a strong EWPT can take place in non-minimal supersymmetric extensions
of the Standard Model \cite{AMNR}.

I thank Michele Maggiore and Luciano Rezzolla for useful discussions.



\end{document}